\documentclass[aps,showpacs,prd,twocolumn]{revtex4}
\usepackage{graphicx}
\usepackage{amssymb}
\usepackage{amsmath}

\newcommand{\be}{\begin{equation}}
\newcommand{\ee}{\end{equation}}
\newcommand{\ben}{\begin{eqnarray}}
\newcommand{\een}{\end{eqnarray}}
\newcommand{\bes}{\begin{subequations}}
\newcommand{\ees}{\end{subequations}}

\begin{document}

\title{Compact vortexlike solutions in generalized Born-Infeld model}
\author{D. Bazeia$^{1}$, E. da Hora$^{1}$ and D. Rubiera-Garcia$^{2}$}
\affiliation{{ $^1$Departamento de F\'isica, Universidade Federal da Para\'\i ba,
58051-900 Jo\~ao Pessoa, Para\'\i ba, Brazil.}\\
$^2${Departamento de F\'isica, Universidad de Oviedo, 33007 Oviedo, Asturias,
Spain.}}

\begin{abstract}
We study vortex-like solutions in a generalized Born-Infeld model. The model
is driven by two distinct parameters, one which deals with the Born-Infeld
term, and the other, which controls the presence of high-order-power term in
the covariant derivative of the Higgs field. We numerically solve the
equations of motion and depict the main vortex features, for several values
of the two parameters of the model. The results indicate the presence
of compact vortex, when the parameter responsible for the high-order-power in the
derivative increases to sufficiently large values.
\end{abstract}

\maketitle


\section{Introduction}

This work deals with the presence of vortices in models described by Abelian gauge field coupled to complex scalar field. In the standard case, the scalar field self interacts via the presence of the Higgs potential, capable of inducing spontaneous symmetry breaking. Moreover, the gauge and scalar dynamics and interactions are controlled by the Maxwell term, and by the usual covariant derivative squared. The system is well known, and capable of  supporting vortex solutions \cite{NO}, which are of general interest to physics \cite{1,2}.

Recently, several authors have studied models with modified dynamics, sometimes known as k-field models. The basic idea is to change the dynamics, allowing for the presence of terms containing higher-order power on the derivative of the field. Motivations to investigate these models appear in cosmology \cite{3}-\cite{8b}, attempting to contribute to understand the present accelerated expansion phase of the universe, and to study issues of strong interaction and other subjects of current interest to high energy physics \cite{8.1}-\cite{18}.

An interesting point of such generalized models is that they may contain defect solutions with compact support \cite{9}-\cite{18}, known as compactons.  If topological defects are seen as field-theoretical realizations of fundamental particles, then compactons are truly localized solutions in the sense that they are absent of the typical infinite solitonic tails, reaching their vacuum values at finite distance, as opposed to conventional solitons. As such infinite tails are more than a modeling nuisance, compactons could be able to model more realistically nature. Compactons first appeared in \cite{19} and have gained many interest in high energy physics and in other recent investigations \cite{20a}-\cite{20d}, including compact bosonic stars \cite{20i}.

In a recent work, some of us studied the presence of compact vortex in the Maxwell-Higgs model, modified to allow for the presence of a term with the fourth-order power in the covariant derivative of the complex scalar field \cite{20f}. The main result suggests that the vortex solutions exhibit compact features when the term introduced to modify the dynamics is important. Motivated by this result, in the present work we go further on the subject, and we investigate rotationally symmetric solutions obtained from a specific k-field theory. This theory is obtained from the Born-Infeld-Higgs Electrodynamics, and it is now endowed with an extra kinecticlike contribution, which is introduced to allow for the presence of a term depending on higher-order power of the covariant derivative of the scalar field. Born-Infeld (BI) theory, originally introduced to deal with the problem of the divergence of the electron self energy in Classical Electrodynamics \cite{BI} is singled out among the class of nonlinear electrodynamics by its special properties under wave propagation \cite{wave} and it currently receives a special attention as it arises in the low-energy limit of the string/D-Branes physics \cite{D-branes1}-\cite{D-branes3}. Moreover in the defects context this action has been found to support global string solutions \cite{20b}.

The aim of this paper is twofold. First to study how the extra kinecticlike term contributes to the compactification of the vortexlike structure present in the model with gauge fields, as such solutions have been already found in some k-field field models \cite{18} as well as in theories with V-shaped potentials (see e.g. \cite{V-shaped}). And second to explore how the BI gauge field affects the existence of solutions which, as shown below, cease to be defined below some critical value of the BI parameter $\beta=\beta_c$, as it has been also suggested to happen in other models \cite{nunez}, indicating that a high enough amount of nonlinearity in the gauge field generically leads to a breakdown in the existence of the solutions within this kind of nonlinear actions.

We organize the paper as follows: in the next Sec.~\ref{general}, we introduce the generalized Born-Infeld-Higgs (gBIH) model and make the basic considerations which allows to get to the equations of motion to describe the vortex configurations. In Sec.~\ref{numerical}, we perform the numerical study and depict the main features of the vortices. Finally, in Sec.~\ref{end} we include comments and conclusions.

\section{Equations of motion}\label{general}

Here we deal with the gBIH model in $(2,1)$ dimensions. The Lagrange density has the form
\begin{equation}\label{model}
\mathcal{L}=\mathcal{L}_{BI}+K\left( X\right) -V\left( \left\vert \phi
\right\vert \right) \text{ .}
\end{equation}%
The (non-zero) diagonal elements of the metric are $(1,-1,-1)$, and the three terms in the above
Lagrange density are given by
\be
\mathcal{L}_{BI}=-\beta ^{2}
\left(\sqrt{1+\frac{1}{2\beta ^{2}}F_{\mu \nu}F^{\mu \nu}}-1
\right) \text{ ,}
\ee
and $K(X)$ is a function of $X$ to be given below, where $X$ stands for
\begin{equation}
X=\left\vert D_{\mu }\phi \right\vert ^{2}\text{ ,}
\end{equation}
and $D_{\mu }\phi$ represents the covariant derivative of the scalar field
\begin{equation}
D_{\mu }\phi =\partial _{\mu }\phi +ieA_{\mu }\phi \text{ ,}
\end{equation}
and $V(|\phi|)$ stands for the potential
\be \label{potential2}
V(|\phi|)=\frac14 \lambda^2(\upsilon^2-|\phi|^2)^2\,.
\ee
In the above expressions, we use $e$ and $\lambda$ as real and positive coupling constants, and $\upsilon$
counts for the symmetry breaking parameter, real and positive.

For simplicity, we introduce the mass scale $M$ and make the scale transformations: $x^{\mu }\rightarrow x^{\mu }/M$, $\phi
\rightarrow M^{1/2}\phi $, $A^{\mu }\rightarrow M^{1/2}A^{\mu }$, $\beta
\rightarrow M^{3/2}\beta $, $e\rightarrow M^{1/2}e$, $\upsilon \rightarrow
M^{1/2}\upsilon $ and $\lambda \rightarrow M^{1/2}\lambda $. In this case, we get $%
X\rightarrow M^{3}X$, $K\rightarrow M^{3}K$ and $\mathcal{L}\rightarrow M^{3}%
\mathcal{L}_{g}$, and the above model is now described by the Lagrange density $\mathcal{L}_g$, which has the same form
\eqref{model}, but now all the coordinates, fields and parameters are dimensionless quantities.

The procedure to search for vortex solutions requires that we investigate the equations of motion. In the present case, they can be written in the form
\begin{equation}
\partial _{\mu }\left( \frac{F^{\mu \alpha }}{\mathcal{R}}\right) =J^{\alpha}\text{ ,}
\end{equation}%
\begin{equation}
K_{X}D_{\mu }D^{\mu }\phi +K_{XX}\partial _{\rho }XD^{\rho }\phi =-\frac{%
\partial V}{\partial \overline{\phi }}\text{ .}
\end{equation}%
Here we use $K_X=dK/dX$, $K_{XX}=d^2K/dX^2$, and
\begin{equation}
J^{\alpha }=ie\left( \overline{\phi }D^{\alpha }\phi -\phi \overline{%
D^{\alpha }\phi }\right) K_{X}\text{ ,}
\end{equation}%
\begin{equation}
\mathcal{R}=\sqrt{1+\frac{1}{2\beta ^{2}}F_{\mu \nu }F^{\mu \nu }}\text{ .}
\end{equation}

The static Gauss Law can be written as
\begin{equation}
\overrightarrow{\nabla }\cdot \left( \frac{\mathbf{E}}{\mathcal{R}}\right)
=-2e^{2}A^{0}\left\vert \phi \right\vert ^{2}K_{X}\text{ ,}
\end{equation}%
where $B=\overrightarrow{\nabla }\times \mathbf{A}$, $\mathbf{E}=-%
\overrightarrow{\nabla }A^{0}$, $\epsilon ^{012}=+1$ and $\cal R$ has now the form
\begin{equation}
\mathcal{R}=\sqrt{1+\frac{1}{\beta ^{2}}\left( B^{2}-\mathbf{E}^{2}\right) }%
\text{ .}
\end{equation}
We fix $A^{0}=0$ as a proper gauge choice and use it from
now on.

As usual, we consider the rotationally symmetric {\it Ansatz} for the fields
\begin{equation}
\phi \left( r,\theta \right) =\upsilon g\left( r\right) e^{in\theta }\text{ ,}
\end{equation}%
\begin{equation}
\mathbf{A}\left( r,\theta \right) =-\frac{\widehat{\theta }}{er}\left(
a\left( r\right) -n\right) \text{ ,}
\end{equation}%
where $r$ and $\theta $ are polar coordinates and $n$ is an integer, the winding number
(vorticity) of the solution. In this case, the rotationally symmetric equations of motion are
\begin{equation}
\frac{d}{dr}\left(\frac{1}{r}\frac{da}{dr}\left( 1+\left( \frac{1}{e\beta r}%
\frac{da}{dr}\right) ^{2}\right) ^{-1/2}\right) =\frac{2e^{2}\upsilon
^{2}g^{2}a}{r}K_{X}\text{ ,}
\end{equation}%
\ben
&&K_{X}
\left(
{\frac{1}{r}\frac{d}{dr}}
\left(
{ r\frac{dg}{dr} }
\right)
-\frac{a^2 g}{r^2}
\right)-{\;\;\;\;\;\;\;\;\;\;\;}\nonumber\\
&&\upsilon ^{2}K_{XX}\frac{dg}{dr}\frac{d}{dr}\left( \left(
\frac{dg}{dr}\right) ^{2}+\frac{g^{2}a^{2}}{r^{2}}\right) =\frac{1}{\upsilon
^{2}}\frac{dV}{dg}\text{ ,}
\een
where $X$ has now the form
\begin{equation}
X=-\left\vert \mathbf{D}\phi \right\vert ^{2}=-\upsilon ^{2}\left( \left(
\frac{dg}{dr}\right) ^{2}+\frac{g^{2}a^{2}}{r^{2}}\right) \text{ .}
\end{equation}

We make the investigation specific, choosing the following forms of $K(X)$ and $V(|\phi|)$
\begin{equation}\label{alpha}
K\left( X\right) =X-\alpha X^{2}\text{ ,}
\end{equation}%
\begin{equation} \label{potential}
V\left( \left\vert \phi \right\vert \right) =\frac{\lambda ^{2}}{4}\left( 1+%
\frac{3}{2}\alpha \right) \left( \upsilon ^{2}-\left\vert \phi \right\vert
^{2}\right) ^{2}\text{ ,}
\end{equation}%
Note that we have introduced a new factor in the potential, depending on $\alpha$,
the parameter which controls the extra term in $K(X),$ responsible for the generalized modification
of the model, as it appears above. The reason for this extra term is explained as follows: when the scalar part in the lagrangian (\ref{model}) is considered, with (\ref{alpha}) for the kinetic term and (\ref{potential2}) for the standard potential, the equations of motion are written as
\begin{equation}
\frac{1}{2} \phi'^2=U(\vert \phi \vert),
\end{equation}
where the new potential $U(\vert \phi \vert )$ can be shown to be given by \cite{20f}

\begin{equation}
U(\vert \phi \vert )=\frac{1}{6} \frac{\sqrt{1+12\alpha V( \vert \phi \vert )}-1}{\alpha},
\end{equation}
which implies that $U(0)=\frac{1}{6} \frac{\sqrt{1+6\alpha}-1}{\alpha}$. For simplicity it is desirable to make the constant $U(0)$ independent of $\alpha$ and choosing for example $U(0)=1/2$ this leads to the modification
\begin{equation}
V(\vert \phi \vert) \rightarrow \left( 1+\frac{3}{2} \alpha \right) V(\vert \phi \vert ),
\end{equation}
which brings the potential (\ref{potential}). As shown in Ref.~\cite{20f} this choice is a good way to investigate the compact profile in theories without the BI term. We take $\alpha$ nonnegative, to avoid problem with the positive definiteness of the energy of field configurations, as the energy density (23) below suggests; see also Ref.~[23].

In this case, for the specific choice above, the rotationally symmetric equations of motion become,
after taking $e=\upsilon =\lambda =1$, for simplicity:
\ben\label{eq1}
&&\frac{d}{dr}\left(\!\frac{1}{r}\frac{da}{dr}\left(1+\left( \frac{1}{\beta r}%
\frac{da}{dr}\right)^{2}\right)^{-\frac12}\right)=\nonumber\\
&&\;\;\;\;\;\;\;\;\;\;\frac{2g^{2}a}{r}\left(
1+2\alpha\left(\left( \frac{dg}{dr}\right) ^{2}+\frac{g^{2}a^{2}}{r^{2}}%
\right) \right),
\een
\ben\label{eq2}
&&\left( 1+2\alpha \left( \left( \frac{dg}{dr}\right) ^{2}+\frac{g^{2}a^{2}}{%
r^{2}}\right) \right) \left( \frac{1}{r}\frac{d}{dr}\left( r\frac{dg}{dr}%
\right) -\frac{a^{2}g}{r^{2}}\right)+\nonumber\\
&&2\alpha \frac{dg}{dr}\frac{d}{dr}%
\left( \left( \frac{dg}{dr}\right)^{2}\!\!+\!\frac{g^{2}a^{2}}{r^{2}}\right)\!=\!g\left(\!1\!+\!\frac{3}{2}\alpha \right) \left( g^{2}\!-1\right).\;\;\;\;\;\;\;\;
\een
These are the equations we have to deal with in the numerical investigation. We recall that the limit $\beta\to\infty$ leads us back to the model studied in Ref.~{\cite{20f}}. Also,
if we further take the limit $\alpha\to0$, we get back to the standard Maxwell-Higgs model. We also note that if we include the Chern-Simons term in the model \eqref{model}, the presence of $\alpha$ generalizes the model studied before in \cite{nunez}, searching for electrically charged vortex solutions in the BI model, in the presence of the Chern-Simons term. However, we will not take this route here, that is, we will not include the Chern-Simons term in this investigation.

To study the presence of vortex solutions to equations \eqref{eq1} and \eqref{eq2}, one needs to know how the fields $a(r)$ and $g(r)$ behave at the origin,
and in the limit $r\to\infty$. Near the origin they should obey
\begin{equation}\label{zero}
g\left( r\rightarrow 0\right) \rightarrow 0\text{ \ \ \ and \ \ \ }a\left(
r\rightarrow 0\right) \rightarrow n=1\text{ ,}
\end{equation}%
where we fix $n=1$, for simplicity.
The asymptotic boundary conditions are obtained from total energy, which is given by
\ben
E=\int \varepsilon \left( r\right) d^{2}r{ ,}
\een
where the energy density has the form
\ben
\varepsilon \left( r\right)& =&\beta ^{2}\left( \sqrt{1+\frac{B^{2}}{\beta ^{2}%
}}-1\right) +\left\vert \mathbf{D}\phi \right\vert ^{2}\nonumber\\
&+&\alpha \left\vert
\mathbf{D}\phi \right\vert ^{4}+\frac{\lambda ^{2}}{4}\left( 1+\frac{3}{2}%
\alpha \right) \left( \upsilon ^{2}-\left\vert \phi \right\vert ^{2}\right)
^{2}{ .}
\een
The energy density must vanish asymptotically, as a condition to make the energy finite. So, we get
\begin{equation}
\underset{r\rightarrow \infty }{\lim }\left\vert \phi \left( r\right)
\right\vert =1 \text{ ,}
\end{equation}%
\begin{equation}
\underset{r\rightarrow \infty }{\lim }\left\vert \mathbf{D}\phi \right\vert
=0\text{ .}
\end{equation}%
From these conditions, using the rotationally symmetric {\it Ansatz}, we get, respectively,
the asymptotic boundary conditions on the profile functions
\begin{equation}\label{infty}
g\left( r\rightarrow \infty \right) \rightarrow 1\text{ \ \ \ and \ \ \ }%
a\left( r\rightarrow \infty \right) \rightarrow 0\text{ .}
\end{equation}

\section{Numerical solutions}
\label{numerical}

Let us now focus our attention to the vortex solutions.
The equations of motion to be considered are \eqref{eq1} and \eqref{eq2}, and
$g$ and $a$ have to obey the conditions \eqref{zero} and \eqref{infty}.

We also plot solutions for the energy density, explicitly written as
\ben
&&\varepsilon \left( r\right) =\beta ^{2}\left( \sqrt{1+\left( \frac{1}{\beta r%
}\frac{da}{dr}\right) ^{2}}-1\right) +\left( \frac{dg}{dr}\right) ^{2}+\frac{%
g^{2}a^{2}}{r^{2}}+\nonumber\\
&&\alpha \left( \left( \frac{dg}{dr}\right) ^{2}+\frac{%
g^{2}a^{2}}{r^{2}}\right) ^{2}+\frac{1}{4}\left( 1+\frac{3}{2}\alpha \right)
\left( 1-g^{2}\right) ^{2}\text{ ,}
\een
and for the  magnetic field
\begin{equation}
B=-\frac{1}{r}\frac{da}{dr}\text{ .}
\end{equation}%
\begin{figure}[t!]
\includegraphics[width=8.0cm,height=4.8cm]{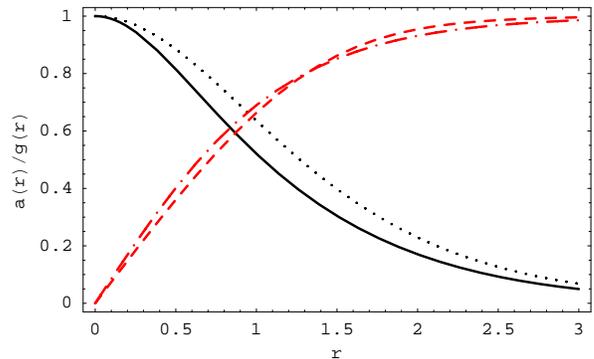}
\caption{We take $\alpha=1$ and $\beta=2$, and we plot $a(r)$ with solid (black) line and $g(r)$ with dashed (red) line, as a function of $r$. We also plot
the standard solution, for $\alpha=0$ and $1/\beta=0$, with $a(r)$ as dotted (black) line and $g(r)$ as dash-dotted (red) line, for comparison.}
\end{figure}
\begin{figure}
\includegraphics[width=8.0cm,height=4.8cm]{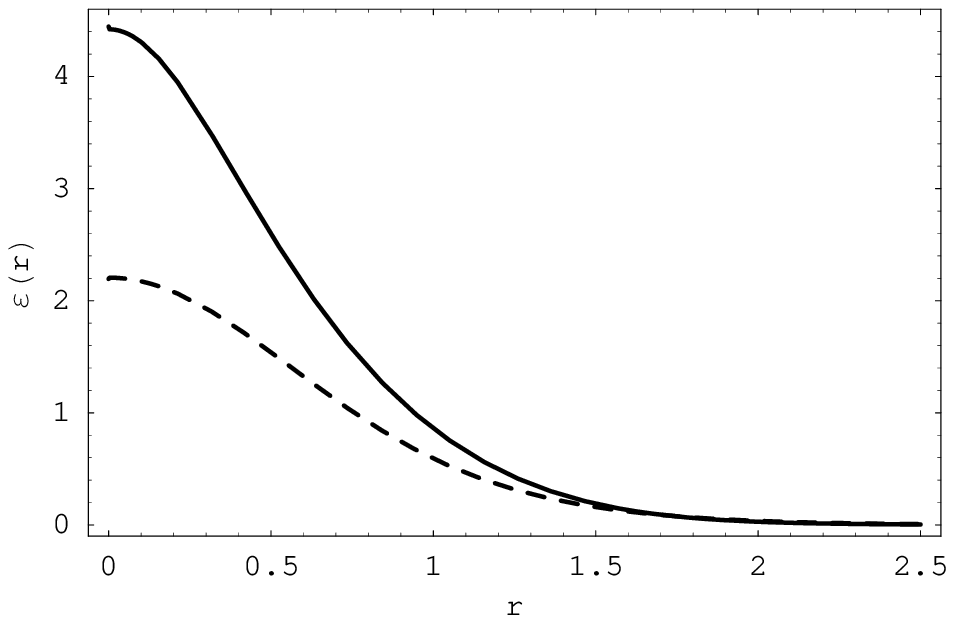}
\includegraphics[width=8.0cm,height=4.8cm]{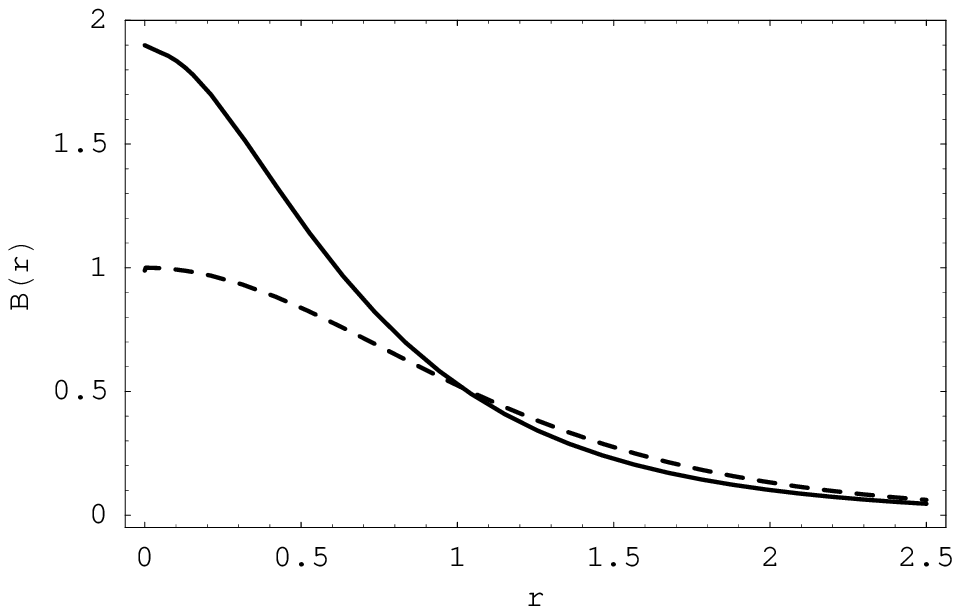}
\caption{We take $\alpha=1$ and $\beta=2$, and we plot $\varepsilon(r)$ (upper panel) and $B(r)$ (lower panel), as a function of $r$. Dashed lines show
the same quantities, but in the standard case, for $\alpha=0$ and $1/\beta=0$.}
\end{figure}

The strategy to numerically solve the equations of motion, given the fact that the boundary conditions do not lie at the same point, is to employ a shooting method, suitably adapted to the system of equations (\ref{eq1})-(\ref{eq2}). Shooting from the center we numerically determine the constants $g'(0)$ and $a''(0)$, up to some predetermined precision, and which are of course $\alpha$ and $\beta$-dependent. This strategy is similar to the one employed, for example, in Ref.\cite{20d}.

\begin{figure}
\includegraphics[width=8.0cm,height=4.8cm]{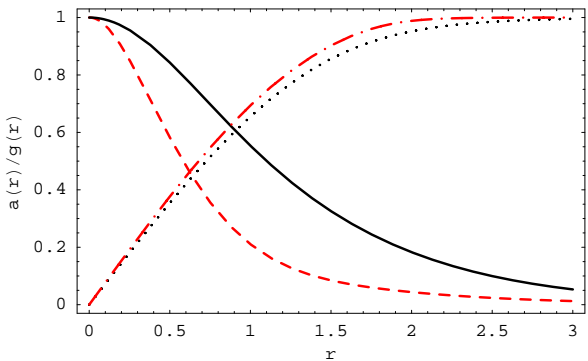}
\caption{We take $\beta=6$, and we plot $a(r)$ and $g(r)$ as a function of $r$, for $\alpha=1$ and  $10$, with solid (black) and dashed (red) lines,
and with dotted (black) and dash-dotted (red) lines, respectively.}
\end{figure}
\begin{figure}
\includegraphics[width=8.0cm,height=4.8cm]{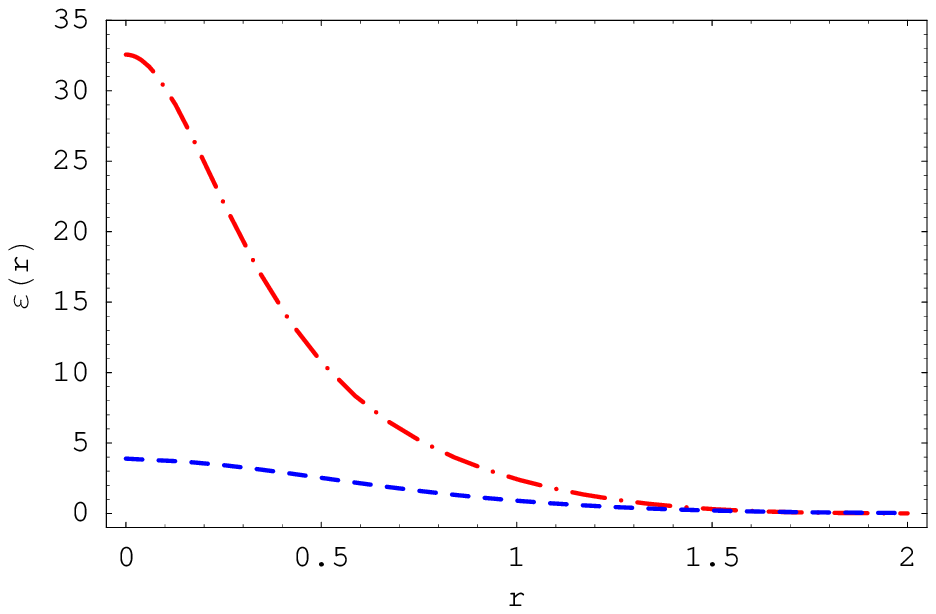}
\\
\includegraphics[width=8.0cm,height=4.8cm]{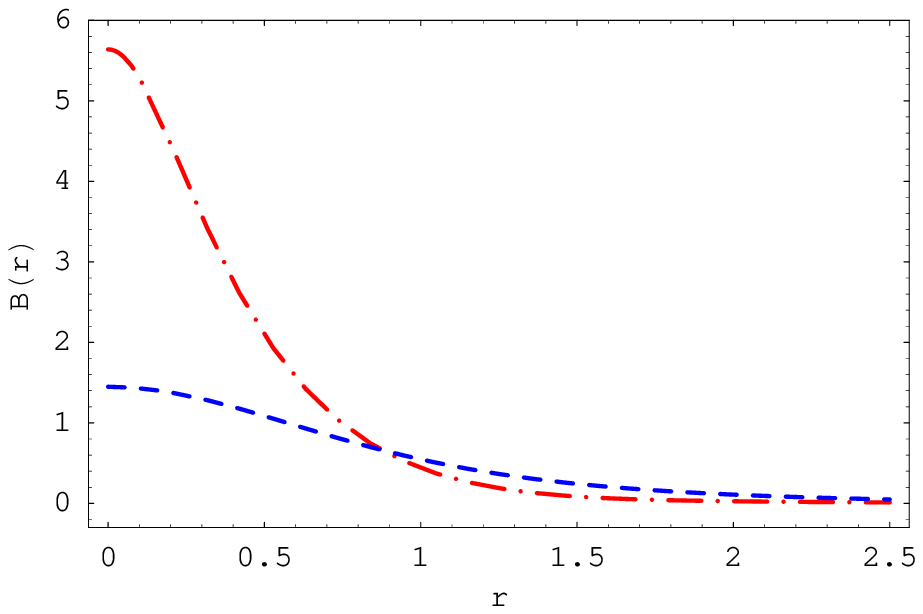}
\caption{We take $\beta=6$, and we plot $\varepsilon(r)$ (upper panel) and $B(r)$ (lower panel), as a function of $r$, for $\alpha=1$ and  $10$,  with
dashed (blue), and dash-dotted (red) lines, respectively.}
\end{figure}

The model to be investigated has two parameters, $\alpha$ and $\beta$, which we have to vary to investigate how the vortexlike solutions depend on them.
The parameter $\beta$ controls the BI term, and for large values of $\beta$, the BI term should behave like the Maxwell term. Also, for $\alpha$ very small, the modification coming from the dynamics of the scalar field should be insignificant; see Eq.~\eqref{alpha}.

In order to illustrate the procedure, let us first consider the case $\alpha=1$ and $\beta=2$. The results for $a(r)$, $g(r)$, $\varepsilon(r)$, and $B(r)$ are all depicted in Figs.~1 and 2. We have investigated the case with $\alpha=1/\beta=0.1$ as well, and it can hardly be distinguished from the standard case, corresponding to $\alpha=1/\beta=0$, which is also plotted in the same Figs.~1 and 2, for comparison. We note from these figures that the values $\alpha=1$ and $\beta=2$ show practically no effect of compactification on the vortexlike solutions.

Let us now concentrate in the case of a large $\beta$, which we choose to be $\beta =6$. In this case, we implement the numerical investigation with a comparison with the previous work \cite{20f}. Thus, we firstly fix $\beta=6$, and we vary $\alpha$, taking the values $\alpha=1$ and $10$, recalling that the degree of importance of the extra
term in $X$ increases with increasing $\alpha$; see Eq.~\eqref{alpha}. The numerical results for the fields $a(r)$ and $g(r)$, the energy density $\varepsilon(r)$ and the magnetic field $B(r)$ are shown in Figs.~3 and 4, and there we see that they behave as they do in the case where the BI term is changed by the standard Maxwell term; compare with results of Ref.~\cite{20f}. In particular, the results depicted in Figs.~3 and 4 clearly show the effects of compactification of the vortexlike structure, as $\alpha$ increases.

Since the numerical study is now under control, we have investigated several distinct cases numerically. Below we comment on some specific values of the parameters $\alpha$ and $\beta$, which we use to illustrate the general situation.
\begin{figure}
\includegraphics[width=8.0cm,height=5.0cm]{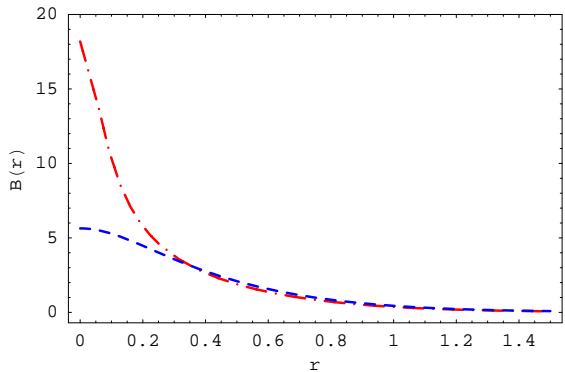}
\caption{We take $\alpha=10$, and we plot $B(r)$ as a function of $r$, for $\beta=6,$ and $4$, with dashed (blue), and dash-dotted (red) lines, respectively.}
\end{figure}

\begin{figure}[tb]
\includegraphics[width=8.0cm,height=4.8cm]{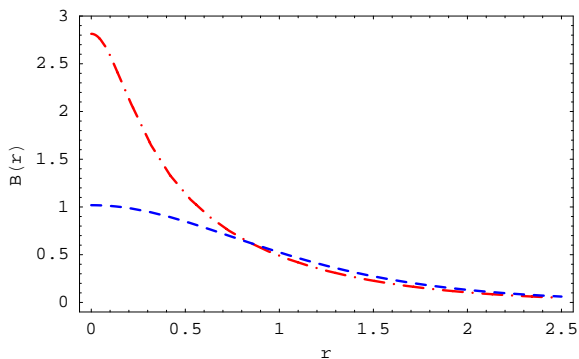}
\caption{We take $\alpha=0$, and we plot $B(r)$ as a function of $r$, for $\beta=5$ and $1$, with dashed (blue), and dash-dotted (red) lines, respectively.}
\end{figure}

It is of current interest to understand if the nonlinearities introduced by the BI term contribute to stress the effects of compactification of the vortexlike structure. To study this issue, let us fix $\alpha$ at a large value, say $\alpha=10$, for which we already know that the vortex tends to have compact profile. Then, we numerically investigate the model for $\beta$ at the values $\beta=6$ and $4$. These results are shown in Fig.~5, where we only plot the magnetic field $B(r)$, which we think illustrate correctly the issue under investigation. Figure 5 shows that the effects of compactification increases as $\beta$ decreases, so the nonlinearities introduced from the BI term for $\beta$ decreasing toward the value $\beta=1$ also contribute to compactification. To check if this effect depends on $\alpha$, let us now investigate the case with $\alpha=0$, for some values of $\beta$. These results are plotted in Fig.~6, for $\beta=5$ and $1$, and they show that the effects of compactification are still there, although they are now much less significant.

It is worth discussing, as a conclusion extracted from the numerical calculations performed, the fact that the solutions, for a given $\alpha$, cease to exist below some critical value $\beta_c(\alpha)$, i.e. they only exist in the range $\beta<\beta_c(\alpha)<\infty$. For example, if we choose the value $\alpha=1$ this critical value is placed around $\beta_c \simeq 1.25$ and for $\alpha=10$ in the range $3.5<\beta_c<4$.  This feature gets manifest, as $\beta$ approaches to $\beta_c$, in a quick increasing of the slope $a_0$ of $a(r)$ near the origin, in order to satisfy the corresponding boundary condition at infinity, in such a way that at the critical value we have $\lim_{\beta \rightarrow \beta_c} a_0(r\simeq 0) \rightarrow \infty$ and the computations break down. This is better seen in the evolution of the magnetic field, as it is defined as the derivative of the $a(r)$ field, which grows without limit as we approach the critical value. This effect seems to be a generic feature of Born-Infeld actions for vortexlike solutions, as can be seen, for example, in \cite{nunez} or \cite{Brihaye2001}, and not something artificially introduced by the numerical procedure employed, as the checking with other numerical recipes (e.g. relaxation methods) may show.

\section{Final comments}
\label{end}

In this work we studied the presence of vortexlike solutions in a BI model coupled to scalar matter field described by generalized dynamics.
The model is controlled by two distinct parameters, $\alpha$, which describes the modification in the dynamics of the matter field, and $\beta$, which drives the BI generalization.

We investigated several regions in parameter space, and we could conclude that for a given $\beta$, the increasing of $\alpha$ increases the effects of compactification of the vortexlike solution. We could also see that for $\alpha$ fixed, the increasing of $1/\beta$ (up to some maximum value $1/\beta_c$) increases the effects of compactification, although in this last case, the effects of compactification are less significant.  However, the general conclusion is that the increasing of $\alpha$ and $1/\beta$ enlarges the effects of compactification, but compactification only occurs at very large values
of $\alpha$. Other quantities such as the energy density and the magnetic field are significantly affected during the evolution with the BI parameter, an effect also observed, for example, in Ref.~\cite{20a}.

The authors would like to thank CAPES and CNPq (Brazil) and FICYT (Spain) for financial support. D.R.-G. would also like to thank the Centro de F\'isica do Porto for their hospitality while doing part of this work and to C. Santos for useful discussions on the numerical issues.


\end{document}